\documentclass[preprint,showpacs,preprintnumbers,amsmath,amssymb,prb]{revtex4}
\usepackage{graphicx}
\usepackage{dcolumn}
\usepackage{bm}
\usepackage{epsfig}
\begin{document}

\preprint{PREPRINT}

\title{Water-like hierarchy of anomalies 
in a continuous spherical shouldered potential}

\author{ Alan Barros de Oliveira}
\affiliation{ Instituto de F\'{\i}sica, Universidade Federal do Rio 
Grande do Sul,  
Caixa Postal 15051, 91501-970, Porto Alegre, Rio Grande do Sul, Brazil.} 
\email{oliveira@if.ufrgs.br}

\author{Giancarlo Franzese}
\affiliation{Departament de F\'{\i}sica Fonamental, Facultat de F\'{\i}sica,
Universitat de Barcelona, Diagonal 647, 08028, Barcelona, Spain.}
\email{gfranzese@ub.edu}

\author{Paulo A. Netz}
\affiliation{ Instituto de Qu\'{\i}mica, Universidade Federal do Rio 
Grande do Sul,  
91501-970, Porto Alegre, Rio Grande do Sul, Brazil.}
\email{netz@iq.ufrgs.br}

\author{ Marcia C. Barbosa}
\affiliation{ Instituto de F\'{\i}sica, Universidade Federal do Rio 
Grande do Sul, 
Caixa Postal 15051, 91501-970, Porto Alegre, Rio Grande do Sul, Brazil .} 
\email{marcia.barbosa@ufrgs.br}

\date{December, 8 2007}

\begin{abstract}

We investigate by molecular dynamics simulations 
a continuous isotropic core-softened potential with attractive well
in three dimensions,
introduced by Franzese [J. Mol. Liq. {\bf 136}, 267 (2007)], 
that displays liquid-liquid
coexistence with a 
critical point and  water-like density anomaly. 
Besides the thermodynamic anomalies, here we find 
diffusion and structural anomalies. 
The anomalies, not observed in the discrete version of 
this model,  occur with the same hierarchy that characterizes water.
We discuss the differences in the anomalous behavior of the continuous
and the discrete model 
in the framework of the excess entropy, calculated
within the pair correlation approximation.

\end{abstract}

\pacs{64.70.Pf, 82.70.Dd, 83.10.Rs, 61.20.Ja}

\maketitle

\section{Introduction}

Experiments for water allow to locate the line of temperatures of maximum
density (TMD)  below which the
density decreases with decreasing temperature 
at constant pressure,  instead of increasing
as in the majority of fluids.
\cite{An76}
Experiments for
Te, \cite{Th76} Ga, Bi, \cite{LosAlamos} 
S, \cite{Sa67,Ke83} 
and Ge$_{15}$Te$_{85}$, \cite{Ts91}  and 
simulations for silica, \cite{An00,Ru06b,Sh02,Po97}
silicon \cite{Sa03} and BeF$_2$, \cite{An00}
show, as well, the same density anomaly.

Water also has dynamic anomalies. Experiments for water show that the diffusion constant, $D$, increases on compression at low
temperature $T$  up to a maximum $D_{\rm max} (T)$ at $P =
P_{D\mathrm{max}}(T)$. The 
behavior of normal liquids, with $D$  decreasing on compression, is
restored in water only at 
high $P$, e.g. for $P>P_{D\mathrm{max}}\approx1.1$ kbar  at 10
$^o$C. \cite{An76} 
Numerical simulations for SPC/E water \cite{spce} recover the experimental
results and 
show that the anomalous behavior of $D$ extends to 
the metastable liquid phase of water at negative pressure,
a region that is difficult to access for experiments.
\cite{Ne01,Er01,Mi06a,Ku06}
In this region the diffusivity $D$ decreases for decreasing $P$  
until it reaches a minimum value 
$D_{\rm min} (T)$ at some pressure $P_{D\mathrm{min}}(T)$, and the
normal behavior, with $D$ increasing for
decreasing $P$, is reestablished only for $P<P_{D\mathrm{min}}(T)$.
\cite{Ne01,Er01,Mi06a,Mu05}

In order to shade some light in the relation between the 
thermodynamic and dynamic anomalies, Errington and Debenedetti investigated the pressure-dependence of the
structure of SPC/E 
water by measuring (i) the tendency of pairs of molecules
to adopt preferential separations, by means of a translational 
order parameter, and (ii) the extend to which a molecule and its four
nearest neighbors assume a tetrahedral arrangement, by means of an  
orientational order parameter. \cite{Er01}
They found a region of the $P$-$T$ phase diagram, called the {\it
structural anomalous region}, where 
both order parameters decrease for increasing density, i.e. the
liquid is more 
disordered at higher density, 
in contrast with the behavior of normal liquids. They
showed that the structural anomalous region
encompasses the region where the diffusion is anomalous, and that the
latter includes the region where the density is anomalous.
Next, Shell, Debenedetti, and Panagiotopoulos used
the two structural order parameters 
to study the 
case of a computer model for silica. 
Both order parameters
demonstrated to be anomalous, 
but the hierarchy of anomalies for silica is 
different compared to water. \cite{Sh02} For silica the 
diffusion anomalous region
is the wider one, containing the structural
anomalous region which in turn
has inside the density anomaly region.

The study of core-softened
potentials generates models that are computationally (sometimes even
analytically) tractable and that may retain some qualitative features
of network forming  fluids such as water and silica.  What they have in common with 
the water potential is the two interaction scales that in 
the case of water are  the van der Waals and the hydrogen bond interaction
distances.   In some cases, 
theoretical 
\cite{Xu06,Ol05,Ol06a,Ol06b,Wi06,Ya05,Ya06,Fr07a} 
and experimental \cite{Sh07} results for core-softened potentials
show  the presence of anomalies. In other cases, in particular 
for the discontinuous shouldered well (DSW) potential, 
no density anomaly is observed. \cite{Fr01,Fr02,Sk04,Ma05}
The DSW potential consists of a hard
core, a square repulsive shoulder and an attractive square well
(dashed line in Fig. \ref{cap:pot}).

In this paper, we investigate the presence of  water-like anomalies in
a continuous shouldered well (CSW) potential 
introduced by Franzese~\cite{Fr07a} (continuous line in Fig. \ref{cap:pot}). 
Although the model is similar to the DSW version, we find density,
diffusion, and 
structural anomalies and we observe that they occur with the same
hierarchy as in water. In order to get
a better understanding of the reasons why the CSW 
potential displays water-like anomalies, while the DSW
one does not, we apply recent ideas
that have emerged on the connection between the anomalies and the excess
entropy.\cite{Ru06b,Mi06a,Mi06b}

Since the model studied here resembles effective potentials for
complex liquids or liquid metals, our results suggest that the
water-like hierarchy of anomalies could be 
experimentally found in systems where pressure-induced bond
shortening, rather than the change in the coordination number, is the
origin of the density anomaly. \cite{Sh07}

The outline of the paper is as follows. 
We present the details of  
the model in Sec. \ref{sec:model}, and the details
of the simulations in Sec. \ref{sec:detsim}.
In Sec. \ref{sec:ressim} we show and discuss
the results obtained from simulations and in Sec. \ref{sec:conc}
we summarize and give the conclusions.

\section{The model} \label{sec:model}
The CSW model \cite{Fr07a} studied here is composed by a system of
identical particles 
interacting through the potential
(Fig. \ref{cap:pot}) defined as 

\begin{equation}
U(r)=
\frac{U_R}
{1+\exp\left[\Delta\left( r-R_R \right)/a\right]}
-
U_A \exp\left[ -\frac{(r-R_A)^2}{ 2\delta_A^2}\right] 
+
\left( \frac{a}{r}\right)^{24},
\label{eq:pot}
\end{equation}

\noindent where 
$U_R$ and $U_A$ are
the energy of the repulsive shoulder and of the attractive well,
respectively,
$a$ is the diameter of particles, 
$R_R$ and $R_A$ are the 
repulsive average radius and the distance of the attractive minimum,
respectively,
$\Delta$ is a parameter related to the slope of the potential at $R_R$;
$\delta_A^2$ is the variance of the Gaussian centered at $R_A$.
The many parameters of this potential 
allow it to assume several forms, similar to the potential proposed by
Cho, \cite{Ch96a} 
ranging from a deep double wells \cite{Ch96a,Ch97a,Ne04}
to repulsive ramp-like potentials, \cite{Ol06a,Ol06b,Ol07} making this equation 
malleable for studying different isotropic models for systems such as
colloids or liquid metals.
Details about the role played by these parameters
may be found in Ref. \cite{Fr07a}.
The set of parameters used here is the same as in Ref. \cite{Fr07a}: 
$U_R/U_A=2,$ $\Delta=15,$ $R_R/a=1.6,$ 
$R_A/a=2,$ $(\delta_A/a)^2=0.1,$
with cutoff $r_c = 3.0 a$. To have a continuous function at the
cutoff distance, a constant $C/U_A=0.208876$ and a linear term $\lambda r/a$,
with $\lambda/a=-0.0673794$, are added. \cite{Fr07a}
The CSW potential
may be interpreted as a continuous
version of the DSW potential
investigated in Refs. \cite{Fr01,Fr02,Ba04,Sk04,Ma05}. The
corresponding DSW
potential is represented by a square shoulder of size $\mathrm{w_R}/a$ and a well
of size $\mathrm{w_A}/a$ and  depth $\mathrm{U_R}+\mathrm{U_A}$ 
illustrated by the dashed line in 
in Fig.  \ref{cap:pot} (the parameters in figure are $\mathrm{w_R}/a=0.612 $, 
$\mathrm{w_A}/a=0.767 $, $\mathrm{U_R}/
\mathrm{U_A}=2.2$).

\section{Details of simulations} \label{sec:detsim}

In order to investigate the interparticle CSW potential 
Eq. (\ref{eq:pot}), we employ a molecular dynamics (MD) 
technique. \cite{Frenkel}
We consider 1000 particles in a cubic box with periodic 
boundary condition.
We perform simulations  in the NVT ensemble with
the Nos\'e-Hoover thermostat \cite{Frenkel} and coupling parameter
equal to 2.
Pressure, temperature, density, and diffusion are calculated in
dimensionless units, 
$P^{*} \equiv P a^3/U_A$,
$T^{*} \equiv k_{B} T/U_A$,
$\rho^{*} \equiv \rho a^3$, and
$D^{*} \equiv D\left(m/a^2 U_{A}\right)^{1/2}$.

Positions, velocities, and configurational energy
of the 
particles are stored for every
$100$ steps, during a total time of $10^6$ steps, given a total
of $10^4$ stored configurations. 
Pressure, diffusion, and
order parameters are calculated over these configurations, 
discarding the first $2\times 10^3$ for equilibration
purposes.  The time step used is $2\times 10^{-3}$ in units of
$(a^2m/U_A)^{1/2}$  
(of the order of $\approx 1.3\times 10^{-12}$s for argon-like atoms 
and $\approx 1.1\times 10^{-12}$s for water-like molecules).

The model displays \cite{Fr07a} a phase diagram with 
two first-order phase transitions
ending in  critical points: one transition between gas and 
low-density liquid (LDL) phases and
another between 
LDL and  high-density liquid (HDL) phase.
The LDL-HDL phase transition is metastable with respect to the crystal
phase.

For temperatures and densities inside the metastable
region the positions, velocities, and configuration energy
are stored for every 10 steps and the 
quantities are calculated with the configurations stored
until the drop in configuration energy and virial, discarding 
the equilibration time. The time before the drop in energy and
virial characterizes the lifetime of the metastable phase.

\section{Results and discussion} \label{sec:ressim}

\subsection{Density anomaly}

In the vicinity of the LDL-HDL
critical point the system shows 
density anomaly. \cite{Fr07a}
The isochores 
have a temperature of minimum pressure
in the interval $0.185<\rho^*<0.215$, 
corresponding to the temperature
of maximum density (TMD) at constant $P$
[Fig. \ref{cap:angellnetzus}(a)].
Inside the TMD line the density decreases as the system 
is cooled at constant $P$. For densities in the range
$0.2<\rho^*<0.215$, the TMD line  
is negatively sloped, while it has positive slope for
$0.185<\rho^*<0.2$.
Our results for the TMD line in the ranges
$0.55\leq T^* \leq 0.61$ and  $0.25\leq P^* \leq 0.30$
[Fig. \ref{cap:angellnetzus}(d)]
compares well, qualitatively, 
 with the experimental data for water, \cite{An76}
showing a negative slope [Fig. \ref{cap:angellnetzus}(b)].
The data in all the supercritical region
[Fig. \ref{cap:angellnetzus}(a)] resemble  
the simulation results for SPC/E water, \cite{Ne01}
with a retracing TMD line [Fig. \ref{cap:angellnetzus}(c)].
Nevertheless, in the present case the TMD line is at $P>0$ for all its
length, while for SPC/E water it extends to negative pressures.

\subsection{Diffusion anomaly} \label{sub:diffanomaly}

The diffusion coefficient is
calculated using the mean-square displacement averaged over different
initial times,

\begin{equation}
\label{eq:Deltar2}
\langle \Delta r(t)^{2} \rangle = \langle [r(t_0+t)-r(t_0)]^2\rangle.
\end{equation}
 
From Eq. (\ref{eq:Deltar2}), the diffusion coefficient may be obtained 
as follows:

\begin{equation}
D=\lim_{t\to\infty}\langle \Delta r(t)^{2} \rangle/6t.
\end{equation}

While for normal fluids diffusivity 
decreases monotonically with increasing density
at constant temperature,
for the model Eq. (\ref{eq:pot}) 
this is the case only for temperatures
$T^{*}>0.64$ (Fig. \ref{cap:diff}).
For $T^{*}< 0.64$,
the diffusion coefficient 
has three regions:
(i) For $\rho<\rho_{D\mathrm{min}}$, $D$
decreases as $\rho$ increases as expected
for normal fluids.
(ii) For $\rho_{D\mathrm{min}} <\rho<\rho_{D\mathrm{max}}$,
$D$ increases with density -- an anomalous behavior. This region is called
the {\it anomalous diffusion region}.
(iii) For $\rho>\rho_{D\mathrm{max}}$, the normal behavior is restored
with $D$ decreasing as density increases.

The line of $P_{D\mathrm{max}}(T)$ 
in the $P$-$T$ phase diagram
[Fig. \ref{cap:angellnetzus}(d)]
is consistent with the 
diffusivity maxima observed in experiments for water
[Fig. \ref{cap:angellnetzus}(b)].
Moreover, both lines of diffusion extrema, $P_{D\mathrm{max}}(T)$ and
$P_{D\mathrm{min}}(T)$ [Fig. \ref{cap:angellnetzus}(a)], 
resemble the results for detailed
models of water \cite{Ne01,Er01,Mi06a} 
[Fig. \ref{cap:angellnetzus}(c)], silica, \cite{Po97,Ru06a,Sh02} and other
isotropic potentials. \cite{Xu06,Ya05,Ya06,Ol06a}
We find that our $D_{\mathrm{min}}$,
like other isotropic models, \cite{Xu06,Ya05,Ya06,Ol06a}
occur at positive $P$.

\subsection{Structural anomaly}

The two quantities used for studying the structural
behavior of the system of particles interacting
through the potential  in Eq. (\ref{eq:pot}) 
are the translational order
parameter, $t,$ and the orientational order parameter,
$Q_6.$  The translational order parameter is defined as
\cite{Er01,Sh02,Er03}

\begin{equation}
t\equiv\int_{0}^{\xi_{c}}|g(\xi)-1|d\xi,
\label{eq:trans}
\end{equation}

\noindent where $\xi\equiv r\rho^{1/3}$ is the  distance
$r$ in units of the
mean interparticle separation
$\rho^{-1/3}$, $\xi_c$ is the cutoff distance set to half of the
simulation box times $\rho^{-1/3}$, as in
Ref. \cite{Ol06b},
$g(\xi)$ is the radial distribution function proportional to the
probability of finding a particle at a (reduced)
distance $\xi$ from a reference particle.
For an ideal gas  $g=1$
and $t=0$. In the crystal phase  $g\ne1$ over long distances and 
$t$ is large. 

For normal fluids, $t$ increases with increasing density. We find this
monotonic behavior only for $T^{*}\geq 1.8$ (Fig. \ref{cap:trans}). For  
$T^{*}<1.8$ we observe that $t$ has a maximum at
$\rho_{t\mathrm{max}}(T)$, decreases for increasing
$\rho>\rho_{t\mathrm{max}}(T)$, reaches a minimum at
$\rho_{t\mathrm{min}}(T)$, and recovers the normal increasing behavior
for $\rho>\rho_{t\mathrm{min}}(T)$ (Fig. \ref{cap:trans}).

For each particle $i= 1, \dots, N$ we
calculate\cite{Er01,Sh02,St83,Er03,To00,Tr00,Hu04}

\begin{equation}
Q_{\ell}^{i}=\left[\frac{4\pi}{2\ell+1}
\sum_{m=-\ell}^{m=\ell}\left| \overline{\left( Y_{\ell m}^{i} \right)}_k
\right|^{2}\right]^{1/2},
\label{eq:Qli}
\end{equation}

\noindent 
with $\ell=6$.
The quantity 
$\overline{\left( Y_{\ell m}^{i} \right)}_k={1 \over k} \sum_{j=1}^k 
Y_{\ell m}(\theta_{ij},\phi_{ij})$ is the average over the 
spherical harmonics $Y_{\ell m}$ 
calculated
over the vectors ${\bf r}_{ij}(\theta_{ij},\phi_{ij})$, with 
$j=1, \dots, k$, 
connecting particle $i$ with its $k$ nearest neighbors $j.$
We use $k=12$ as in Ref. \cite{Ol06b}.
The local orientational order \cite{Ya06,Ol06b} of the whole system is
calculated as 
the average over the $N$ particles,

\begin{equation}
Q_6=\frac{1}{N}\sum_{i=1}^{N}Q_{6}^{i}.
\label{eq:Q6}
\end{equation}

\noindent 
For a crystal, $Q_6$ is large, while
$Q_{6}^{\mathrm{ig}}=1/\sqrt{k}$ for an ideal gas.

Our study reveals an anomalous behavior also for the orientational
order parameter, $Q_6$. For normal fluid, $Q_6$ increases monotonically
with increasing density. Instead, for the present model 
we find that $Q_6$ is non monotonic, 
displaying a maximum at $\rho_{Q\mathrm{max}}$, for all the
considered temperatures
(Fig. \ref{cap:orient}).  

Since $\rho_{Q\mathrm{max}}$ lies between the densities which bound
the extrema in $t$, $\rho_{t\mathrm{max}}$ and $\rho_{t\mathrm{min}}$,  
we call {\it structural anomalous region} the interval between
$\rho_{Q\mathrm{max}}$ and $\rho_{t\mathrm{min}}$. In this region 
\emph{both} parameters $t$ and $Q_6$ are anomalous, and the liquid
becomes less ordered with increasing density, in contrast with the
behavior of normal liquids.

\subsection{The hierarchy of anomalies and order map}

The relation between the several anomalies presented for the CSW
potential is illustrated in Fig. \ref{cap:allrhotemp}.
The TMD line lies between the
diffusivity extrema (DE) lines, that are included within the 
structural anomalous region bounded by the curves of maxima of $Q_6$
and minima of $t$.
The hierarchy of anomalies found here is the same reported for the
SPC/E water \cite{Er01} and other two-scales potentials.\cite{Xu06,Ol06b,Ya06} 

The {\it order map} in the $t-Q_6$ plane 
(Fig. \ref{cap:ordermap}) resembles the one observed for SPC/E water,
\cite{Er01} silica, \cite{Sh02} and other two-scale potentials
\cite{Ya06,Ol06b} 
since it has an inaccessible region.
Differently from water,
and similar to other two-scale potentials, \cite{Ya06,Ol06b}
the translational and orientational order parameters 
are not coupled into the structural anomaly region.
Indeed, for the densities belonging to the structural anomalous region,
$t$ and
$Q_6$ map onto a two dimensional region (Fig. \ref{cap:ordermap}), in
opposition 
to the water case, in which they map onto 
a single line. \cite{Er01}

\subsection{The Widom line and the  minima in $t$}

The so called Widom line is defined as the locus of the maximum
of response functions close to the critical point at the fluid phase.
Here we calculate the isothermal compressibility (Fig. \ref{cap:ptall}),
\begin{equation}
\kappa_T = - \left( \frac{\partial \ln V}{\partial P} \right)_T.
\end{equation}

As expected, the Widom line (diamond symbols in Fig. \ref{cap:ptall})
is a continuation of the liquid-liquid coexistence line (crosses
in Fig. \ref{cap:ptall}). Our data show that
the Widom line
coincides with the minima in the translational order parameter $t$, at
least in the vicinity of the liquid-liquid critical point.
This is consistent with the observation that
compressibility, structure factor, $g(r)$ and
the parameter $t$ are related.
In particular, we find that $\kappa_T$ has its maximum where the
translational order is minimum. Hence, the largest variation of volume
at constant $T$ for increasing $P$ occurs at the pressure where the
behavior of the structural parameter $t$ changes from anomalous to
normal.
To our knowledge these are the first data 
showing this relation between $\kappa_T$ and $t$.

\subsection{Excess entropy and anomalies}

Why the DSW 
potential  has no water-like anomalies and
its continuous counterpart, the CSW potential, does? We
can gain some understanding by analyzing the density
dependence of the excess entropy. It has been shown\cite{Ru06b}
that the behavior of the excess entropy as a function of density 
is linked to anomalies in density, diffusion and structure.   

Extrema in $(\partial S / \partial \rho)$ correspond to an
thermal expansion coefficient equal to zero and therefore
to density extrema.\cite{Ru06b}
The thermodynamic condition that gives rise to the 
density anomaly can be written as
$\Sigma_{\mathrm{ex}}\equiv \left(\partial
s^{\mathrm{ex}}/\partial \ln \rho\right)_T>1$, as shown by 
Errington \emph{et al.} in Ref.\cite{Er06}. 
Here $s^{\mathrm{ex}}$ is the excess entropy,
defined as $s^{\mathrm{ex}}=s-s^{\rm ig}$, i.e. 
the difference between the entropy $s$
of a real fluid and the entropy $s^{\rm ig}$
of an ideal gas at the same $T$ and $\rho$, due to the correlations
between the position of the particles of the real fluid. 

Based on the proposition of Rosenfeld\cite{Ro99} that the logarithm
of the diffusion coefficient $D$ is 
proportional to the excess entropy, a non-monotonic behavior 
in the excess entropy would imply a non-monotonic behavior in $D$.
 Following the empirical Rosenfeld's parameterization, \cite{Ro99} 
Errington \emph{et al.}  \cite{Er06} have also observed that the
diffusion anomaly can be predicted by $\Sigma_{\mathrm{ex}} > 0.42$.  

Finally, they argue that $\Sigma_{\mathrm{ex}} > 0$ 
is a good estimate for determining the region where structural
anomaly occurs, because for normal fluids the excess
entropy decreases for increasing density at constant temperature. \cite{Er06}
Recent works have explored the new possibilities one
can achieve through the connections between excess entropy and
structure, not only in the matter of isotropic fluids 
\cite{Mi06b,So07,Ru06b} but also
in water,\cite{Sc00a,Mi06a} silica, \cite{Ru06b}
and BeF$_2$. \cite{Ag07,Agunpub}

To calculate the excess entropy, one should count all the accessible
configurations for a real fluid and compare with the ideal gas
entropy. This calculation is not straightforward  and
can approximated by
\begin{equation}
s_{2}=-2\pi\rho\int\left[g(r)\ln g(r)-g(r)+1\right]r^2 dr,
\end{equation}
since $s_2$ is the dominant contribution to excess 
entropy\cite{Gr52,Ra71,Ba89} and it is proved to be 
between 85$\%$ and 95$\%$ of the total excess entropy in Lennard-Jones
systems.\cite{Ba89,Cha06}
The two-body contribution $s_2$ depends only on 
the radial distribution function $g(r)$ and the density, giving
a direct connection between structure and thermodynamics. 
The excess entropy and the translational order parameter are linked 
because both depend on the deviation of $g(r)$ from unity. The
relation between excess entropy and the orientational order parameter,
however, depend on the symmetries of the structures in the high and
low density limits.  Indeed, it turns out that 
the extent to which the pair correlation entropy is sensitive
to orientational order determines the behavior of the structurally anomalous
region.\cite{Ru06b}

However, as shown in Ref. \cite{Er06} 
$s_2$ overestimates the excess entropy,
$s^{\mathrm{ex}}< s_2$, and $\Sigma_{\mathrm{ex}} < \Sigma_{2} 
\equiv \left(\partial s_{2}/\partial \ln \rho\right)_T$.
Therefore, the inequalities $\Sigma_{2} > 0$, $\Sigma_{2} > 0.42$,
and $\Sigma_{2} > 1$, overestimate the region of structural, diffusion
and density anomalies, respectively. Nevertheless, 
$\Sigma_{2}$ has been shown to give estimates of the anomalous
regions in qualitative agreement with the estimates based on
$\Sigma_{\mathrm{ex}}$. \cite{Er06}

In order to compute $\Sigma_{2}$ for the DSW
potential, we simulate 500 particles inside a cubic box
with periodic boundary conditions. The particles interact
through the DSW potential (dashed line in Fig. \ref{cap:pot}),
with parameters $\mathrm{w_R}/a=0.612$, $\mathrm{w_A}/a=0.767$, and 
$\mathrm{U_R} / \mathrm{U_A}=2.2$. The equilibration
and production times are 350 and 650, respectively, in reduced units.
To achieve the desired temperature, we 
rescale the velocities until the equilibration. 
At equilibrium, we
simulate the system in the $NVE$ ensemble.

From the analysis of $s_2$ and $\Sigma_{2}$ for the DSW potential
[Fig. \ref{cap:s2-todos}(a) and Fig. \ref{cap:s2-todos}(b)],
we observe that $\Sigma_{2}$ has a maximum, not monotonic with
$T$, that is always smaller than 1. 
Therefore, $\Sigma_{\mathrm{ex}}<1$ and no density anomaly is
expected, in agreement with Franzese \emph{et al.}.  \cite{Fr01}

We find densities where the DSW potential has 
$\Sigma_{2}>0.42$, for $0.60 \leq T^* \leq 0.75$,  and
$\Sigma_{2}>0$, for all the four temperatures studied here
[Fig. \ref{cap:s2-todos}(b)]. These values apparently suggest the presence of
diffusion and structural anomaly, respectively.
However, since $\Sigma_{2}>\Sigma_{\mathrm{ex}}$ 
and 
the Rosenfeld criteria holds with
$30\%$ uncertainty, \cite{Er06} our results for $\Sigma_{2}$
in Fig. \ref{cap:s2-todos}(b) give no final answer about the presence
of diffusion anomaly.
A more detailed study, beyond the goal of this work, on
diffusion and structural anomaly for the DSW 
potential would be necessary to clarify these points.

Our results on $s_2$ and $\Sigma_{2}$ for the CSW
potential are consistent with our analysis of the anomalies described
in the previous sections
[Fig. \ref{cap:s2-todos}(c) and Fig. \ref{cap:s2-todos}(d)].
We observe that $\Sigma_{2}$ displays a maximum that increases for
decreasing temperature.
For the four studied temperatures, we find that $\Sigma_{2}>0.42$
(hence, greater than zero),
suggesting the occurrence of both diffusion and structural anomalies.
Since we find a range of densities where $\Sigma_{2}>1$ only 
for  $T^*\leq 0.75$, the result suggests the presence of density
anomaly only below this temperature. 
This is qualitatively consistent with the
water-like hierarchy of anomalies that we have reported in the
previous sections.
However, the comparison between 
Fig. \ref{cap:s2-todos}(d) and Fig. \ref{cap:allrhotemp} shows that the 
criteria based on $\Sigma_{2}$ overestimate 
the temperatures and densities where the anomalies appear.
We interpret this discrepancy as the effect of using $\Sigma_{2}$,
instead of $\Sigma_{\mathrm{ex}}$, in the criteria for the anomalies,
with  $\Sigma_{2}>\Sigma_{\mathrm{ex}}$.

Therefore, the excess entropy analysis is a useful
tool to distinguish between the DSW and the CSW potential.
In the discontinuous case the increase of the excess entropy with
density at constant $T$ is not enough to give rise to the density
anomaly. In the continuous case, instead, its increase is large enough
at low $T$. For the diffusion and structural anomalies, the analysis
based on the approximate expression $s_2$ is less clear. However, the
comparison with the direct calculations of the anomalies in the CSW
case suggests that only for 
the CSW potential the increase of excess entropy is enough to reach
the regimes with diffusion and structural anomalies at the
considered temperatures.

From Fig. \ref{cap:s2-todos} one observes that for the CSW potential
$s_2$ is more sensitive to density variations than for the DSW
potential when $\rho^* <0.24$, i. e. when the average interparticle
distance $r/a\equiv (\rho^*)^{-1/3} > 1.612$. 
This is consistent with the fact that the DSW
potential is constant for $1.612\leq r/a < 2.379$, corresponding to
average density $\rho^* \leq 0.24$ and  $\rho^* > 0.07$, respectively. 
Therefore, for the DSW potential in this regime of density one can
expect a small variation of $g(r)$ and, as a consequence, of $s_2$.
Instead, the CSW potential for $r/a > 1.612$ ($\rho^* <0.24$) has a
large variation, equal to 100\% of its attractive energy depth,
implying a sensible variation in $g(r)$ and $s_2$.

It is worth noticing that, recently, Netz \emph{et al.} \cite{Ne06} have
argued that, in potentials with two characteristic scales,
the region of density anomaly can be observed only if it appears at 
a temperature such that $k_BT$ is larger than 
the discontinuity in the interaction
energy (e.g., $k_BT>U_A+U_R$ in the DSW in
Fig. \ref{cap:pot}). 
This is consistent with our results for the CSW potential, 
showing the density anomaly at $k_BT=0.65~U_A>0$, because the CSW
potential has no discontinuity in the interaction energy.
On the other hand, 
for the DSW potential with discontinuity
$U_A+U_R=3.2~U_A$, since no density anomaly is 
found for $k_BT>3.2~U_A$, 
this criterion excludes that the anomaly could be found at lower temperatures.

\section{Summary and conclusions} \label{sec:conc}

We studied a three dimensional system 
of particles interacting  through a continuous isotropic interparticle
pair potential recently proposed by Franzese. \cite{Fr07a} 
The potential has a hard-core,
a repulsive shoulder and an attractive well.
This continuous shouldered well (CSW) potential 
resembles the discontinuous shouldered well (DSW) potential
studied in Ref. \cite{Fr01} and shows new features that are absent in
the DSW potential.

\begin{table}[ht]
\caption{\label{criticalpoints} 
Critical temperatures $T^*_{C_1}$ and $T^*_{C_2}$, pressures
$P^*_{C_1}$ and $P^*_{C_2}$, and densities $\rho^*_{C_1}$ and $\rho^*_{C_2}$,
for the gas-liquid critical point $C_1$ and the liquid-liquid critical
point $C_2$, 
for the CSW potential. The quantities are expressed in dimensionless units.}
\begin{ruledtabular}
\begin{tabular}{lccc}
Critical point & $T^*$ & $P^*$ &  $\rho^*$\\
\hline
Gas-Liquid $C_1$  &$0.95\pm 0.06$ &$0.019\pm 0.008$ &$0.08\pm 0.03$\\
Liquid-Liquid  $C_2$ &$0.49\pm0.01$ &$0.285\pm0.007$  &$0.247\pm0.008$\\
\end{tabular}
\end{ruledtabular}
\end{table}

As its discontinuous counterpart, the CSW
potential displays a
phase diagram with a gas-liquid phase transition ending in a critical
point, and a LDL-HDL phase transition ending in a liquid-liquid critical
point. Table \ref{criticalpoints} shows the values of the critical
parameters for the two critical points.
The LDL-HDL phase transition is
metastable with respect to the crystal phase. \cite{Fr07a} 
In the $P$-$T$ phase diagram the LDL-HDL phase transition line has
positive slope, as in the discontinuous model, \cite{Fr01} 
and in the ramp-potential. \cite{Ja01a} This feature
suggests that these models are describing systems different from
water, because for water and water-like models the slope of the
liquid-liquid phase transition is expected to be negative.
\cite{
Po92,Br05,Fr03,Vo01b} 

In contrast with its discontinuous counterpart, the CSW 
potential has density anomaly (Fig.\ref{cap:ptall}). \cite{Fr07a} 
The TMD line bends toward the LDL-HDL critical point at high $P,$ as
also seen for other isotropic potentials. \cite{Ja01a}

We find that the CSW potential has also diffusion and structural
anomalies (Fig. \ref{cap:ptall}), as water and silica. 
Since water \cite{Er01} and silica \cite{Sh02} have different order of
these anomalies, we investigate the cascade of anomalies for the CSW
potential. We find that for the CSW model the structural anomalous
region encompasses 
the diffusion anomaly region, inside which the density anomaly region
is observed, as in water. This is consistent with previous analysis of
other two-scale isotropic
potentials \cite{Ya05,Ol06b}.  

The loci of structural anomalies and the two phase transition lines in
the $P$-$T$ phase diagram display a positive slope.
The TMD line has positive slope
at low $P$ and negative slope at high $P$. 
By decreasing $T$, we find that three quantities
-- the minima of the structural order parameter, the Widom line, and
the maxima of diffusivity -- approach the LDL-HDL critical
point value, 
suggesting that the high-$T$ behavior of these quantities
could give an indirect indication of the location of the LDL-HDL
critical point in real systems, if it is present. In particular, close
to the LDL-HDL critical point, the
Widom line coincides with the minima of the translational structural
order parameter.

To understand the difference between the DSW and the CSW 
potential we perform an excess entropy analysis, approximating
the excess entropy by the pair correlation entropy $s_2$.
Comparison of this analysis with the direct
calculation of the anomalies suggests that the
approximate excess entropy $s_2$ satisfies criteria with 
empirical thresholds higher then those predicted for
the exact excess entropy $s^{\mathrm{ex}}$.

Interestingly, the excess entropy analysis for the DSW and the CSW
potential emphasizes a relevant difference
for the appearance of the anomalies: 
the DSW potential is constant in the density regime of the
anomalies (approximately for $0.17<\rho^*<0.25$, i. e. average
interparticle distance $1.6<r/a<1.8$), while the CSW
potential has $r$-dependent soft repulsion. 
This observation could clarify also why potentials,
such as the ramp in Ref.\cite{Ja01a}, show water-like anomalies,
because they have appreciable distance dependence (soft repulsion) for
interparticle separation between the repulsive and attractive length. 

Our work confirms that water-like anomalies can be present also in 
systems that, unlike water, have
no directional bonds, as discussed in previous works.
\cite{Ol06a,Ol06b,Ol05,Wi06,Xu06,Ya05,Ya06,Ba04}
The CSW potential, and other two-scales isotropic models
\cite{Ol06a,Ol06b,Ol05,Wi06,Xu06,Ya05,Ya06,Ba04,Ja01a},  
have features that are different from water, such as 
the positive slope in the $P$-$T$ phase diagram of the 
LDL-HDL phase transition, or the locus of diffusion minima 
at positive pressure, instead of negative. \cite{Ne01} 
These differences emphasize 
the possible existence of liquids with anomalies induced by bond
shortening at high $P$, rather than the change in the coordination
number as in water or silica.
This conclusion is
consistent with recent results on metallic glasses showing
polyamorphism. \cite{Sh07}

We conclude by remarking that, although a similar cascade of anomalies
was observed in other two-scales potentials,  \cite{Xu06,Ol06b,Ya06}
the CSW potential is the first with a strong shoulder that
displays these anomalies. This feature make the CSW potential suitable
to study liquid
metals, such as Cs, Mg, Sr, Ba, \cite{Wa00} and colloids,
\cite{colloid2} where effective interactions with large shoulder
are derived in some cases. By varying the parameters of
Eq. (\ref{eq:pot}), the 
potential changes from deep double wells to a repulsive ramp, and by
varying the power of the core of Eq. (\ref{eq:pot}) -- here
equal to 24 to mimic a hard-core -- the stiffness of the core changes,
making the Eq. (\ref{eq:pot}) an interesting functional form to model
and interpolate pressure-dependent effective interactions.

\subsection*{Acknowledgements}

We thank for financial support 
the Brazilian science agencies CNPq, CAPES, and FINEP  
and the Spanish
Ministerio de Educaci\'on y Ciencia for the International Cooperation
IRCR Grant No. PHB2004-0057-PC.
G. F. acknowledges financial support from the Spanish
Ministerio de Educaci\'on y Ciencia within the 
Programa Ram\'on y Cajal and Grant No. FIS2004-03454.

\bibliographystyle{aip}
\bibliography{Biblioteca}

\pagebreak

\section*{Captions to the figures}

Fig. 1: (Color online) Interaction potentials studied in this work. 
Dashed line represents  the discontinuous shouldered well (DSW)
potential. \cite{Fr01,Fr02,Sk04,Ma05}
Continuous line represents the continuous shouldered well
(CSW) potential introduced in Ref.~\cite{Fr07a}.
The parameters are explained in the text.

Fig. 2: (Color online) (a) Pressure-temperature ($P$-$T$) diagram for the CSW model.
Lines correspond to isochores with,
from bottom to top,
$\rho^{*}=$ 0.175, 0.18, 0.185, 0.19, 0.2, 0.21, 0.215, 0.22,
0.23, 0.24, 0.25, 0.26, 0.27, 0.28, 0.29, 0.3, 0.31, and 0.32.
Triangles represent spinodal lines for the LDL-HDL phase transition,
with LDL at low $P$ and HDL at high $P$. We estimate the LDL-HDL
critical point where the spinodal lines converge (large filled circle).
Line with crosses represents our estimate of the
liquid-liquid coexistence line.
The TMD (bold continuous) line bends toward the LDL-HDL critical
point at high $P$.
Dashed lines bound the region where
the diffusion
anomaly occurs (see section  \ref{sub:diffanomaly}).
(b) Experimental data for water anomalies adapted from 
Angell \emph{et al.}. \cite{An76} 
Circles denote the line 
of temperatures of maximum density (TMD) 
at constant $P$.
Squares mark
where the diffusion has a maximum value with
increasing $P$ at constant $T$, $P_{D\mathrm{max}}(T)$.
(c) Simulation data for SPC/E water adapted from 
Netz \emph{et al.}. \cite{Ne01} Squares mark
$P_{D\mathrm{max}}$
where the diffusion has a maximum value, $D_{\rm max}$,
at constant $T$, and diamonds
mark $P_{D\mathrm{min}}$
for the local minima,  
$D_{\rm min}$.
Circles locate the TMD line.
(d) Zoomed region from panel (a), showing good qualitative agreement
between our simulations and the experiments.

Fig. 3: (Color online) The diffusion coefficient against the density for several isotherms.
For the range of densities bracketed within the dashed lines,
the particles move faster under compression 
for temperatures lower than 0.64. This is the opposite
behavior which one expects for normal fluids. 
Dashed lines are guides for the eyes connecting $\rho_{D_{min}}$ and
$\rho_{D_{max}}$.

Fig. 4: (Color online) The translational order parameter as a function of density.
While for normal fluids compression leads to increase the translational 
order parameter, for the model of Eq. (\ref{eq:pot}) this is the case
only for high  
temperatures ($T^{*}\geq 1.8$).
Dashed lines bound the region where $t$
behaves anomalously.

Fig. 5: (Color online) The orientational order parameter against density. 
We observe that $Q_6$ has a maximum at $\rho_{\rm Q max}$, 
meaning that
$Q_6$ decreases under compression for some range of densities. 
The $Q_6$ maxima lie between the extrema points of the 
translational order parameter, $t$. 
Dashed line marks the location of maximum $Q_6$.

Fig. 6: (Color online) Temperature-density plane containing all the anomalies
found for the CSW potential. 
The TMD line bounds the innermost region with the density anomaly behavior.
This region is surrounded by the $D$ extrema lines, that encompass the region
with diffusion anomaly.
The out-most anomalous region, including the first two, is 
between curves B and A, where the system exhibits
an anomalous behavior in structure as shown by the order parameters
$t$ and $Q_6$. The curve C marks the maxima in $t$ occurring where
$Q_6$ has a normal behavior.

Fig. 7: (Color online) The $t-Q_6$ plane or order map. The arrows indicate the direction
of increasing density. Each line correspond to an isotherm and from
top to bottom they are
$T^{*}=$ 0.49, 0.55, 0.60, 0.62, 0.65, 0.70, 0.75, 0.80, 0.90, 1.0,
1.3, 1.5, 1.7, and 1.8. 
By increasing the density, at low $\rho$ both order parameters
increase (normal fluid behavior), then at intermediate $\rho$ they both
decrease (structural  anomaly region), then at higher $\rho$ the
orientational $Q_6$ decreases, while the translational $t$ increases.
As in the case of SPC/E water, silica and other two-scales potentials 
the region with high $Q_6$ and low $t$ is inaccessible. The
inaccessible region is limited by a straight line  
$t_{\rm min}=a+b Q_6$ with $a=-2.86\pm0.02$ and $b=12.8\pm0.1$.

Fig. 8: (Color online) Pressure-temperature phase diagram merging all the results
found for the CSW potential.
The meaning of the lines
is described in the legend, where DE stands for diffusivity extrema,
LL for liquid-liquid, and LG for liquid-gas. 
See the text for more details.

Fig. 9: (Color online) (a) Pair contribution of excess entropy, $s_2$,
for the DSW potential (dashed line in Fig. \ref{cap:pot}) against
density at constant $T$. Circles
are simulated data and lines are fifth order polynomial fit from data.
(b) $\Sigma_{2}= \left(\partial s_2 / \partial \ln \rho \right)_T$
is shown for DSW potential. Panels (c) and (d) show the results for
the CSW model.
Horizontal lines mark the threshold value for 
anomaly in density $\rho$, diffusion $D$, and structure, 
as explained in the text.
Solid, dotted, dashed, and dotted-dashed lines
correspond to temperatures $T^* =$ $0.55$, $0.60$,
$0.75$, and $0.90$ in all panels.

\pagebreak

\begin{figure}[ht]
\begin{center}\includegraphics[clip=true,scale=0.6]{POTENTIAL.eps}
\end{center}
\caption{
\label{cap:pot}}
\end{figure}

\pagebreak

\begin{figure}[ht]
\begin{center}

\includegraphics[%
  scale=0.30]{DENSITY_ANOMALY.eps}\includegraphics[%
  scale=0.30]{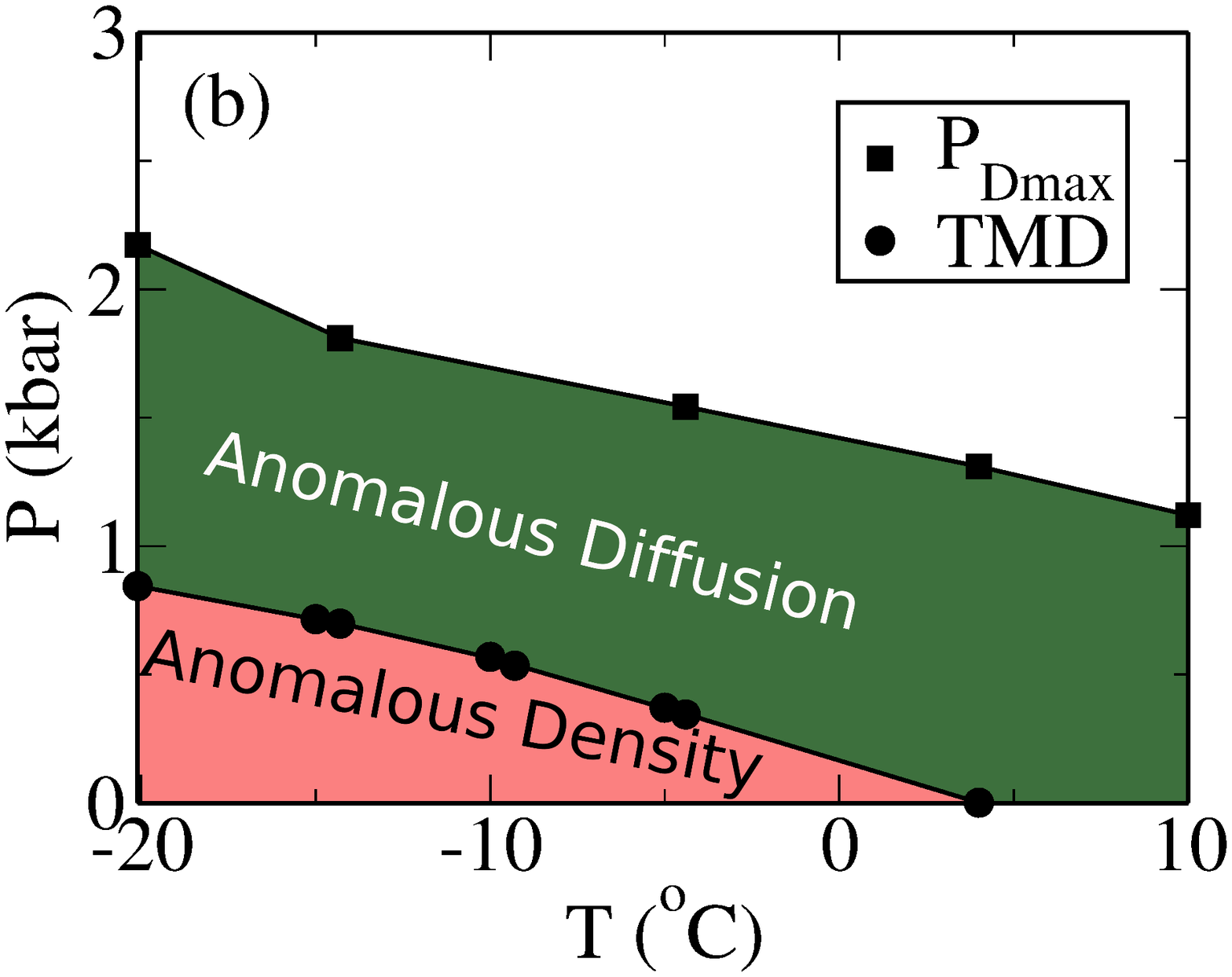}

\includegraphics[%
  scale=0.30]{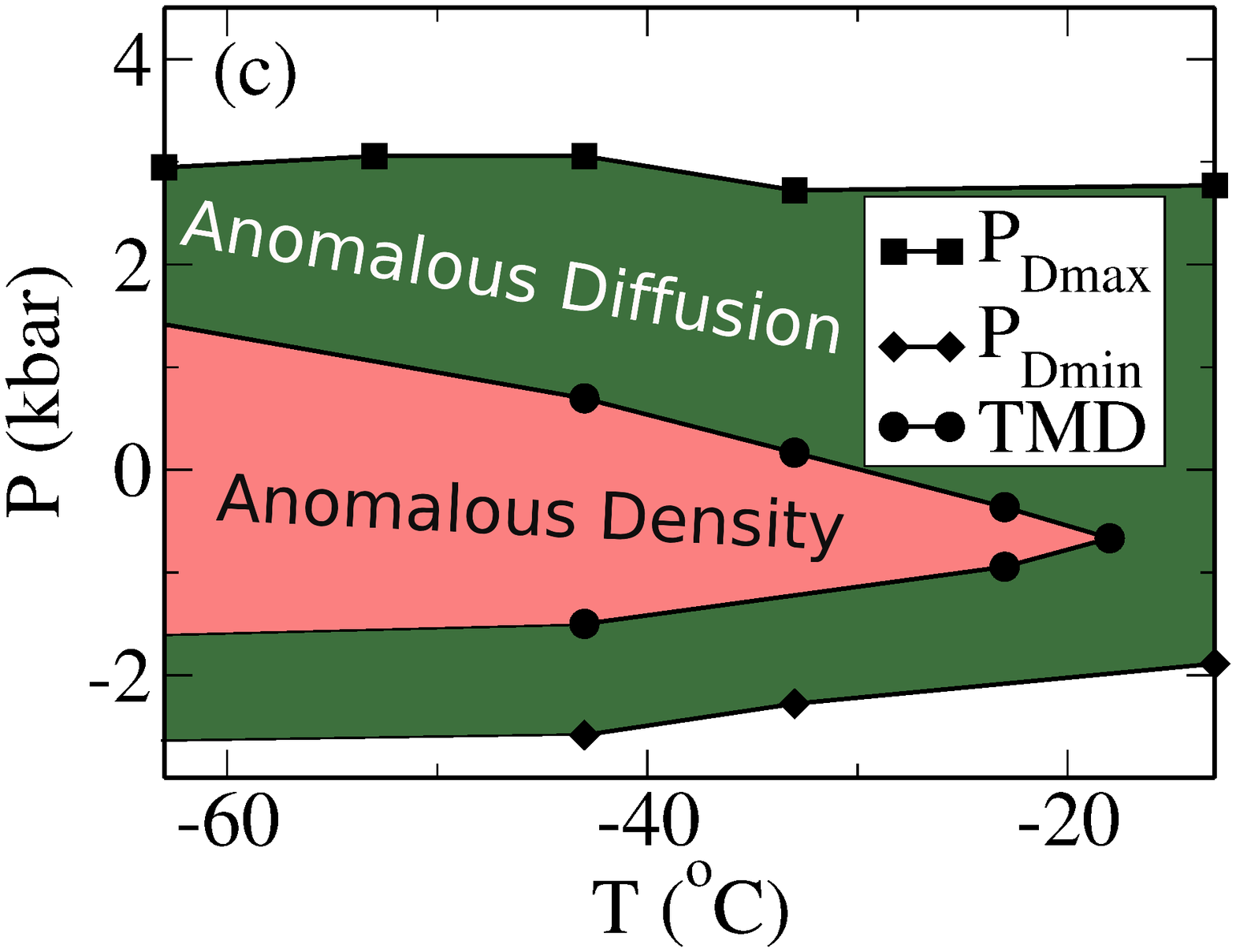}\includegraphics[%
  scale=0.30]{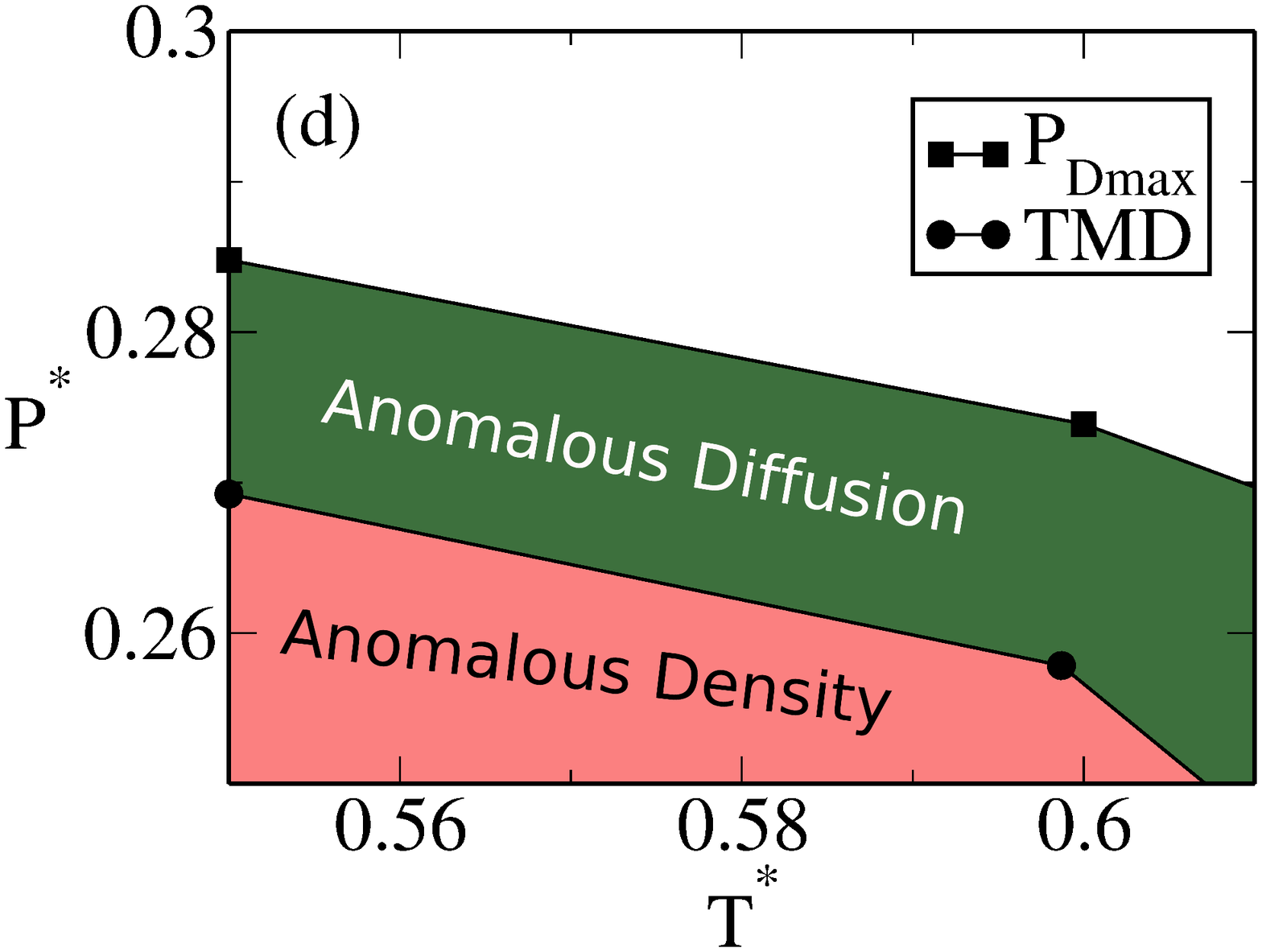}

\end{center}
\caption{
\label{cap:angellnetzus}}
\end{figure}

\pagebreak

\begin{figure}[ht]
\begin{center}\includegraphics[clip=true,scale=0.5]{DIFF.eps}
\end{center}
\caption{ 
\label{cap:diff}}
\end{figure}

\pagebreak

\begin{figure}[ht]
\begin{center}\includegraphics[clip=true,scale=0.5]{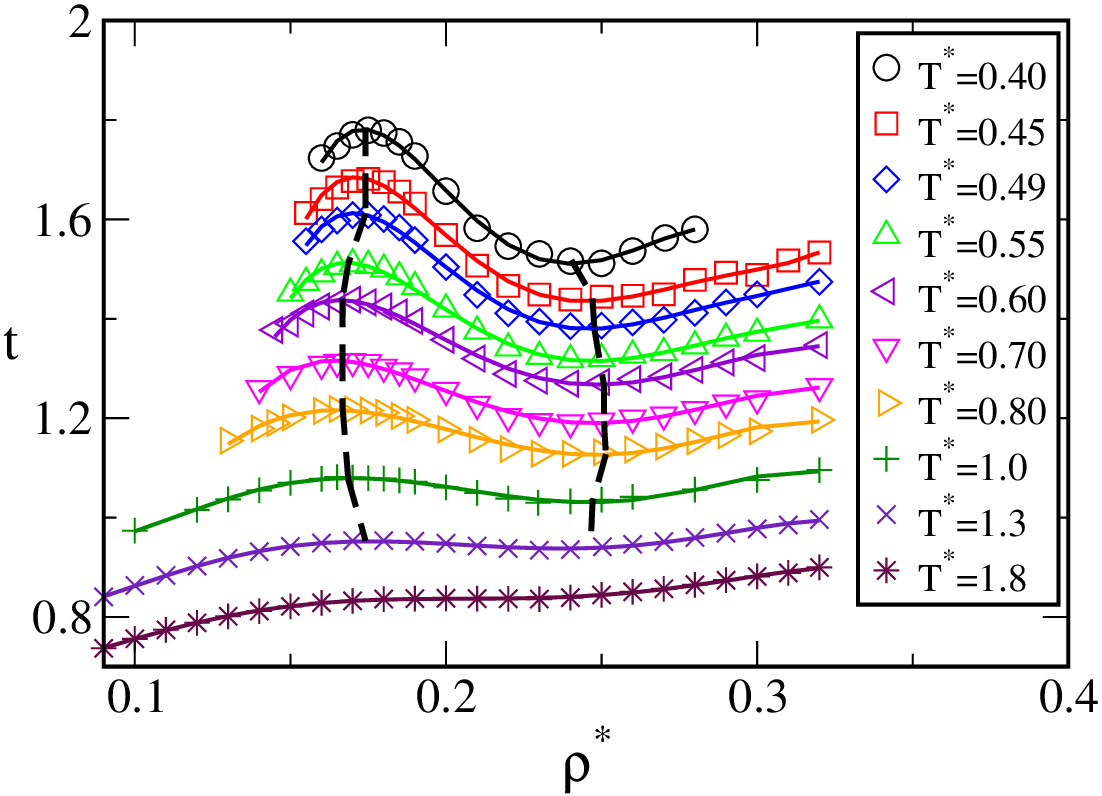}
\end{center}
\caption{
\label{cap:trans}}
\end{figure}

\pagebreak

\begin{figure}[ht]
\begin{center}\includegraphics[clip=true,scale=0.5]{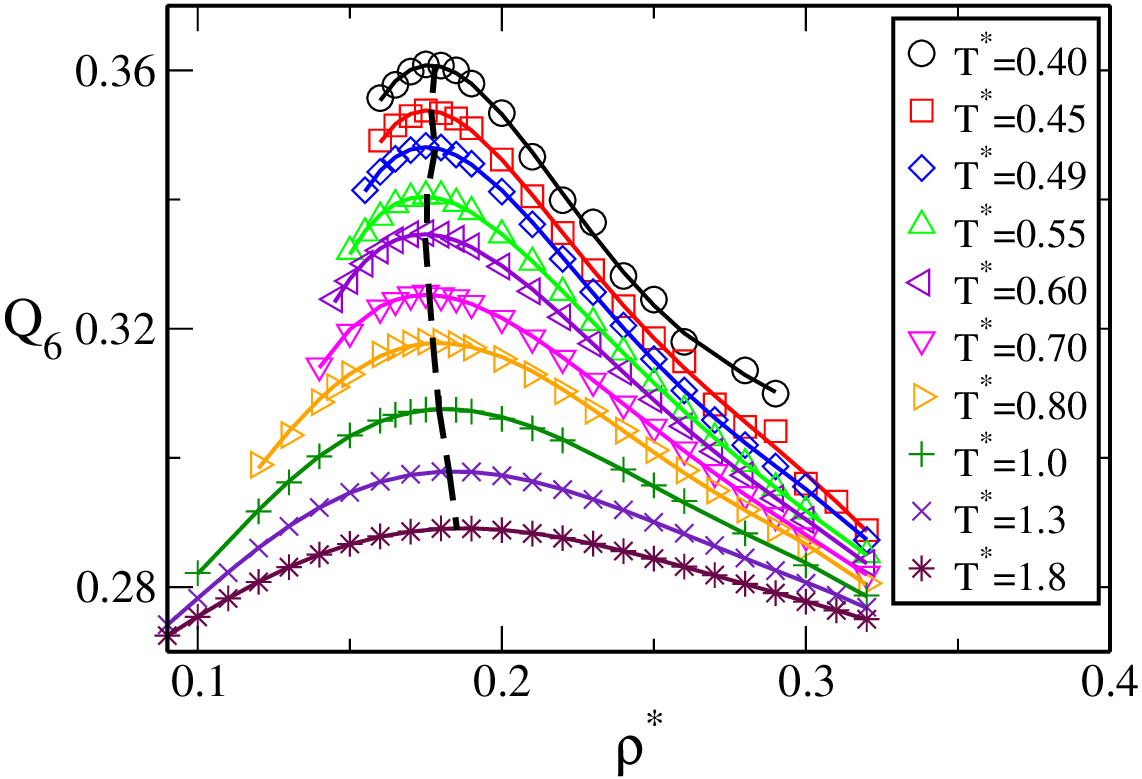}
\end{center}
\caption{
\label{cap:orient}}
\end{figure}

\pagebreak

\begin{figure}[ht]
\begin{center}\includegraphics[clip=true,scale=0.65]{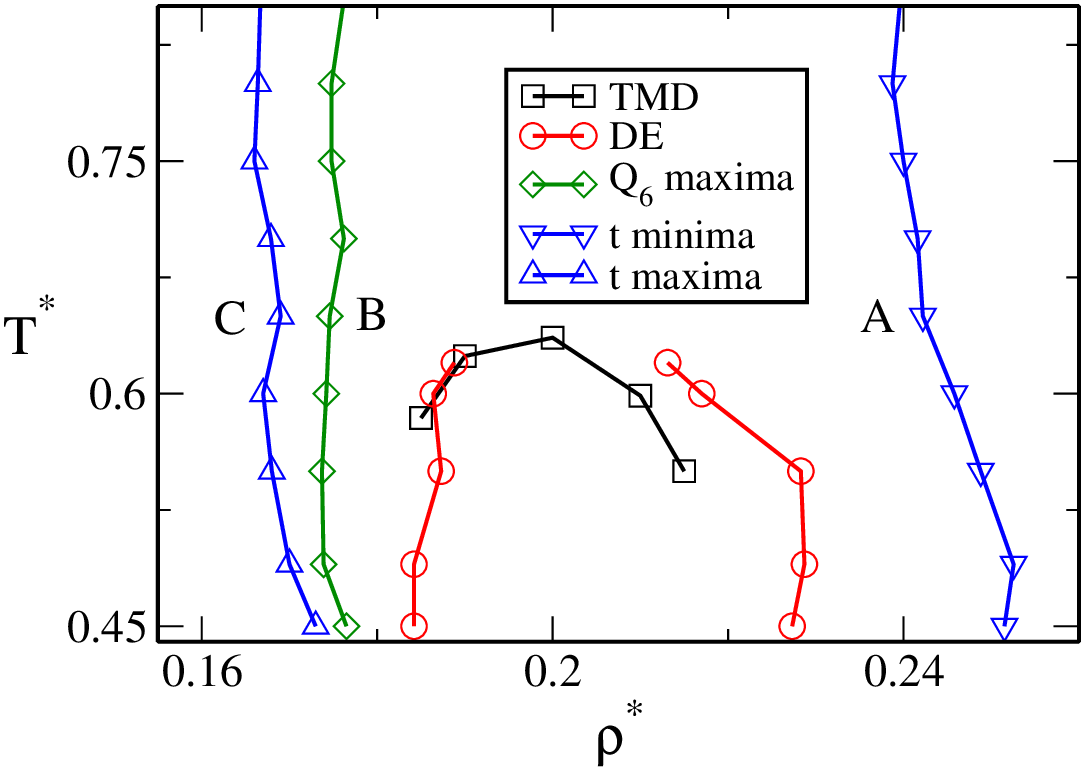}
\end{center}
\caption{
\label{cap:allrhotemp}}
\end{figure}

\pagebreak

\begin{figure}[ht]
\begin{center}\includegraphics[clip=true,scale=0.5]{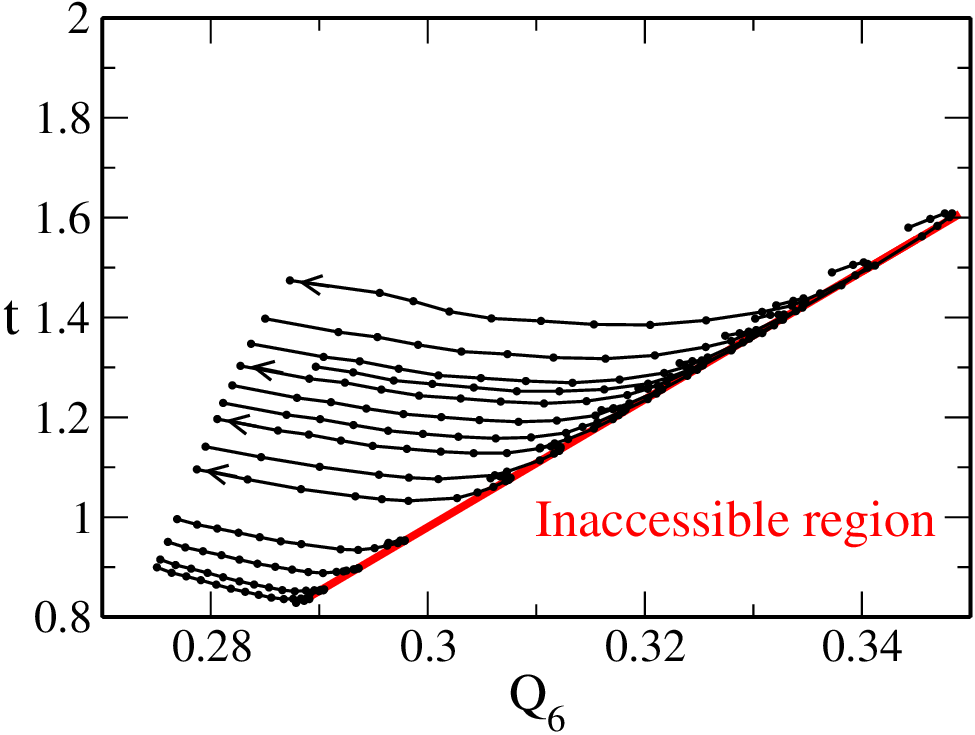}
\end{center}
\caption{
\label{cap:ordermap}}
\end{figure}

\pagebreak

\begin{figure}[ht]
\begin{center}\includegraphics[clip=true,scale=0.65]{PTALL2.eps}
\end{center}
\caption{
\label{cap:ptall}}
\end{figure}

\pagebreak

\begin{figure}[ht]
\begin{center}\includegraphics[clip=true,scale=0.65]{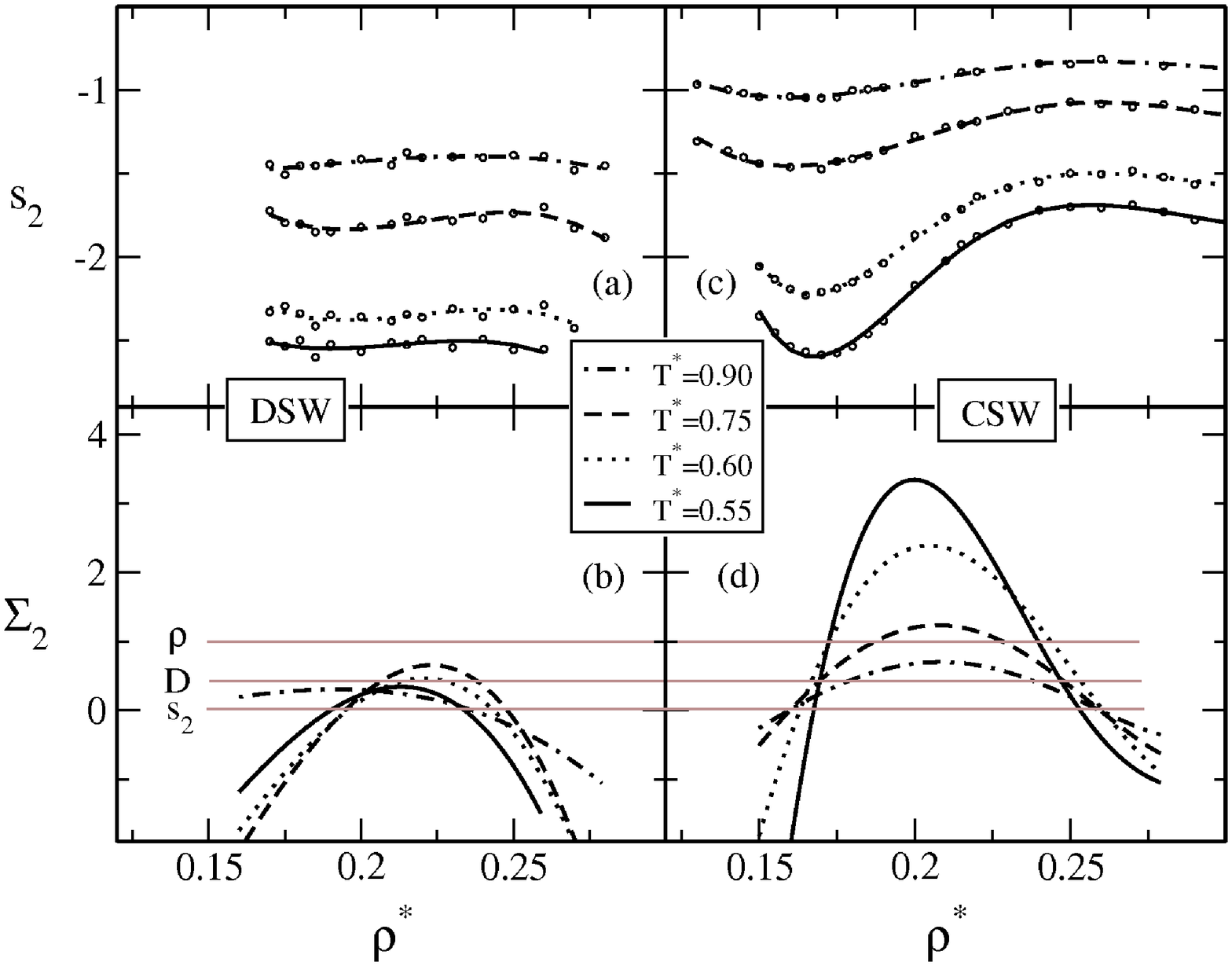}
\end{center}
\caption{ 
\label{cap:s2-todos}}
\end{figure}

\end{document}